# DYNAMIC GLUON CONFINEMENT IN HIGH ENERGY PROCESSES WITHIN EFFECTIVE QCD FIELD THEORY

K. Geiger

*CERN TH-Division, CH-1211 Geneva 23, Switzerland*

## Abstract

An effective Lagrangian approach to describe the dynamics of confinement and symmetry breaking in the process of quark-gluon to hadron conversion is proposed. The deconfined quark and gluon degrees of freedom of the perturbative QCD vacuum are coupled to color neutral condensate fields representing the non-perturbative vacuum with broken scale and chiral symmetry. As a first application the evolution of gluons emitted by a fragmenting high energy $q\bar{q}$ pair from the perturbative to the non-perturbative regime with confinement is studied. For reasonable parameter choice the solution of the equations of motion leads to flux tube configurations with a string tension $t \simeq 1$ GeV/fm.

e-mail: klaus@surya11.cern.ch

The transformation of color charged quarks and gluons in high energy QCD processes into colorless hadrons is commonly believed to be a local phenomenon. The universality of parton fragmentation observed in hard processes as $e^+e^-$ annihilation, deep inelastic $ep$ scattering, Drell Yan, etc., is strongly supported by experimental observations. Thus, a consistent description of the local hadronization mechanism must be independent of the details of the partons prehistory and should apply also to hadron-hadron, hadron-nucleus, or nucleus-nucleus collisions.

To date most of the theoretical tools to study properties of QCD are inadequate to describe the dynamics of the transformation from partonic to hadronic degrees of freedom: Perturbative techniques are limited to the deconfined, short distance regime of high energy partons [1], QCD sumrules [2] and effective low energy models [3] are restricted to the long distance domain of hadrons, and lattice QCD [4] lacks the capability of dynamical calculations concerning the quark-gluon to hadron conversion. On the other hand, phenomenological approaches to parton fragmentation [5], are mostly based on hadronization models with adhoc prescriptions to simulate hadron formation from parton decays.

In this letter I will propose a universal approach to the dynamic transition *between* partons and hadrons based on an effective QCD field theory description. The key element is to project out the *relevant degrees of freedom* for each kinematic regime and to embody them in an effective QCD Lagrangian which recovers QCD with its scale and chiral symmetry properties at high momentum transfer, but yields at low energies the formation of symmetry breaking gluon and quark condensates including excitations that represent the physical hadrons. I will first formulate the general field theoretical framework. Subsequently I shall demonstrate its applicability to the dynamics of parton-hadron conversion by studying exemplarily the evolution of gluons produced by a fragmenting quark-antiquark pair. The aim of this letter is to give a lucid presentation of the conceptual framework, details of the work reported here will be given elsewhere [6].

The effective QCD Lagrangian is represented in the form

$$\mathcal{L} := \mathcal{L}[\psi, A] + \mathcal{L}[\chi, U] + \mathcal{L}[\psi, A, \chi]. \tag{1}$$

Here

$$\mathcal{L}[\psi, A] = -\frac{1}{4} F_{\mu\nu,a} F_a^{\mu\nu} + \overline{\psi}_i \left[ i\gamma_\mu \partial^\mu - g_s \gamma_\mu A_a^\mu T_a \right] \psi_i + \mathcal{L}_{gauge} + \mathcal{L}_{ghost} \tag{2}$$

is the QCD Lagrangian in terms of the quark ($\psi_i$) and gluon fields ($A_a$) and $F_a^{\mu\nu} = \partial^\mu A_a^\nu - \partial^\nu A_a^\mu + g_s f_{abc} A_b^\mu A_c^\nu$. The effective Lagrangian [7]

$$\mathcal{L}[\chi, U] = \frac{1}{2}(\partial_\mu \chi)(\partial^\mu \chi) + \frac{1}{4} Tr\left[(\partial_\mu U)(\partial^\mu U^\dagger)\right] - c\left(\frac{\chi}{\chi_0}\right)^3 Tr\left[m_q(U + U^\dagger)\right]$$
$$- b\left[\frac{1}{4}\chi_0^4 + \chi^4 \ln\left(\frac{\chi}{e^{1/4}\chi_0}\right)\right] - \frac{1}{2}\left(\frac{\chi}{\chi_0}\right)^4 m_0^2 \phi_0^2 , \qquad (3)$$

introduces a scalar gluon condensate field $\chi$ and a meson field $U = f_\pi \exp\left(i\sum_{j=0}^{8} \lambda_j \phi_j / f_\pi\right)$ for the nine pseudoscalar meson fields $\phi_j$ ($f_\pi = 93$ MeV), with non-vanishing vacuum expectation values $\chi_0 = \langle 0|\chi|0\rangle \neq 0$ and $U_0 = c\langle 0|U + U^\dagger|0\rangle \neq 0$ that explicitly break scale, respectively chiral symmetry. In (3), $c$ is a constant, $m_q = \text{diag}(m_u, m_d, m_s)$ and $b$ is related to the conventional bag constant $B$ by $B = b\chi_0^4/4$. The key ingredient in (1) is the connection between the scale and chiral symmetric, short distance regime of colored fluctuations and the world of colorless hadrons with broken symmetries. It is mediated by the coupling between the fundamental quark and gluon degress and the effective fields $\chi$ and $U$ through coupling functions $g(\chi)$ and $\xi(\chi)$,

$$\mathcal{L}[\psi, A, \chi] = \frac{\xi(\chi)}{4} F_{\mu\nu,a} F_a^{\mu\nu} - \overline{\psi}_i g(\chi) \psi_i . \qquad (4)$$

The couplings are chosen [6] in the spirit of the Friedberg-Lee model [8] as $g(\chi) = g_0 \xi/(\xi - 1)$ and $\xi(\chi) = \theta(\chi) v^3 (3v - 4)$ with $v = \chi/\chi_0$ which generate absolute confinement of quarks and gluons [9], here intimately connected with the formation of the gluon condensate $\chi_0$ and the $q\bar{q}$ condensate $U_0$. The non-linear couplings have the properties that $\xi = g = 0$ for $\chi = 0$ at short distances, but in the long range regime $\xi \to 1$, $g \to \infty$ as $\chi \to \chi_0$.

The specific form of $\Delta\mathcal{L} \equiv \mathcal{L}[\chi, U] + \mathcal{L}[\psi, A, \chi]$ is motivated by the *dual QCD vacuum picture* [8]: High momentum, short distance quark-gluon fluctuations (the perturbative vacuum) are embedded in a confining background field $\chi$ (the non-perturbative vacuum), in which by definition the low momentum, long range fluctuations are absorbed. In analogy to a thermodynamic system in equilibrium with a heat bath there can be a net flow of energy between the system and the heat bath environment such that the *bare* energy of the system is not conserved. However the *free* energy of the system, here high momentum quarks and gluons, is constant [10]. It corresponds to the conserved physical energy momentum tensor $\theta_{\mu\nu}$ with the well known trace anomaly [11]:

$$\theta_{\mu\mu} = (1 + \gamma_m) \sum_q m_q \bar{q} q + \frac{\beta(\alpha_s)}{4\alpha_s} F_{\mu\nu} F^{\mu\nu} , \qquad (5)$$

where $\gamma_m$ is the anomalous mass dimension and $\beta$ the QCD beta function. This anomaly constraint is modeled [7] by the third and fourth term in (3) with the correspondence $U_0 = \langle 0|\bar{q}q|0\rangle$ and $-B\chi_0^4 = \langle 0|(\beta/2g_s)F_{\mu\nu}F^{\mu\nu}|0\rangle$.

The following remarks concerning the effective QCD Lagrangian (1) are important: First, at short distances the exact QCD Lagrangian is recovered, since $\chi = U = 0$ and $\xi = g = 0$, whereas the long distance QCD properties emerge as $\chi/\chi_0, U/U_0 \rightarrow 1$ [7] and no colored quanta survive. Second, the problem of double counting degrees of freedom has to be carefully inspected. Although it does not arise in one-loop calculations (to which I will restrict here), processes with e.g. two-gluon exchange could also be contained in the exchange of a color singlet $\chi$-quantum. A minimal possibility to avoid this problem is a rigid separation of high and low momentum modes, in a way that it is exclusive but complementary. Third, $\mathcal{L}[\chi, U]$ embodies the correct QCD scaling properties. It accounts for the anomaly (5) without which Poincaré invariance would be broken, and the mass of the proton would come out wrong, since $2m_p^2 = \langle p|\theta_\mu^\mu|p\rangle$. Fourth, there is no need for renormalization of $\mathcal{L}[\chi, U]$ since $\chi$ and $U$ are already interpreted as effective, composite fields with loop corrections implicitly included.

The effective QCD field theory defined by (1) is readily applicable to describe the dynamic evolution from perturbative to non-perturbative vacuum in high energy processes. In the following I shall consider as an example the fragmentation of a $q\bar{q}$ jet system with its emitted bremsstrahlung gluons and describe the evolution of the system as it converts from the parton phase to the hadronic phase. A time-like virtual photon in an $e^+e^-$ annihilation event with large invariant mass $Q^2 \gg \Lambda^2$ is assumed to produce a $q\bar{q}$ pair which initiates a cascade of sequential gluon emissions. The early stage is characterized by emission of "hot" gluons far off mass shell in the perturbative vacuum. Subsequent gluon branchings yield "cooler" gluons with successively smaller virtualities, until they are within $Q_0^2$, where $Q_0$ is of the order of $m_\chi \equiv 4b\chi_0^2 \simeq 1$ GeV. At this point condensation sets in, or loosely speaking, the "cool" gluons are eaten up by the color neutral scalar condensate field $\chi$ representing the non-perturbative vacuum. Restricting the analysis to the gluonic sector, the equations of motions that derive from (1) reduce to:

$$\partial_\mu \left[\kappa(\chi)F_a^{\mu\nu}\right] = -g_s\kappa(\chi)f_{abc}A_{\mu,b}F_c^{\mu\nu} \tag{6}$$

$$\partial_\mu \partial^\mu \chi = -\frac{\partial V(\chi)}{\partial \chi} - \frac{\partial \kappa(\chi)}{\partial \chi} F_{\mu\nu,a} F_a^{\mu\nu} \tag{7}$$

$$\partial_\mu \partial^\mu U = -4c \left(\frac{\chi}{\chi_0}\right)^3 Tr[m_q], \tag{8}$$

where $\kappa = 1 - \xi$ and $V(\chi)$ is the potential evident from (3). The key problem is the first equation, since the gluon fields drive the dynamics of the $\chi$-field which in turn feeds back via $\kappa(\chi)$. Note that the $U$-field does not couple directly to the gluon field. The procedure in the following is therefore to 'solve' eq. (6) as a function of $\kappa$ and then to insert the solution into (7), so that one is left (aside from the simple third equation) with a single non-linear equation for $\chi$. As a simplification I shall work in the *mean field approximation*, i.e. separating off the quantum fluctuations of the $\chi$-field, $\chi = \bar{\chi} + \hat{\chi}$, where $\bar{\chi}$ is a c-number and $\hat{\chi}$ a quantum operator (similarly for $U$).

Solving the equation of motion (6) for the gluon field is equivalent to the calculation of the complete Greens function with an arbitrary number of gluons, a formidable task indeed. Instead, I will restrict to evaluate the 2-point Greens fuction only (Fig. 1), i.e. the full gluon propagator which includes both the gluon self-interaction through real and virtual emission and absorption, and the effective interaction with the confining background field $\chi$ [6]. In the framework of "jet calculus" [12], this gluon propagator describes how a gluon, produced with an invariant mass $Q^2$, evolves in the variable $x$ and the virtuality $k^2$ (or transverse momentum $k_\perp^2$) through these interactions. In one-loop approximation it is obtained by calculating the corresponding single loop cut diagrams. In the present case, one has in addition to the usual gluon branching and fusion processes, $g \to gg$ and $gg \to g$, contributions from energy transfer and two gluon annihilation processes $g \to g\chi$ and $gg \to \chi$, respectively. This is illustrated in Fig. 1.

The determining equation for the propagator corresponds an evolution equation for the gluon distribution, which is defined [13] as average number of gluons at 'light cone time' $r^+ = 0$ in the multi-gluon state $|P\rangle$ with light cone fractions $x \equiv k^+/P^+$ in a range $dx$ and transverse momenta in a range $d^2 k_\perp$ with respect to the jet axis:

$$x\, g(x, k_\perp^2) = \frac{1}{P^+} \int \frac{dr^- d^2 r_\perp}{(2\pi)^3} e^{-i(k^+ r^- - \vec{k}_\perp \cdot \vec{r}_\perp)} \langle P| F_a^{+\mu}(0, r^-, \vec{r}_\perp)\, F_{a,\mu}{}^+(0, 0, \vec{0}_\perp) |P\rangle \tag{9}$$

Performing a standard Mellin transformation to the *moment representation* $g(\omega, k_\perp^2) :=$

$\int_0^1 dx\, x^\omega g(x, k_\perp^2)$, the evolution equation reads:

$$k_\perp^2 \frac{\partial}{\partial k_\perp^2} g(\omega, k_\perp^2) = \gamma(\omega, k_\perp^2)\, g(\omega, k_\perp^2), \tag{10}$$

where $\gamma(\omega, k_\perp^2)$ plays the role of a generalized *anomalous dimension* ($\lambda_\chi \equiv \xi(\chi)^2/4\pi$),

$$\begin{aligned}\gamma(\omega, k_\perp^2) &= \frac{\alpha_s(k_\perp^2)}{2\pi}\, \theta(k_\perp^2 - Q_0^2) \left[A_{g\to gg}(\omega) - \left(\frac{\Lambda}{k_\perp^2}\right) A_{gg\to g}(\omega)\right] \\ &\quad + \frac{\lambda_\chi(k_\perp^2)}{2\pi} \left[A_{g\to g\chi}(\omega) - \left(\frac{\Lambda}{k_\perp^2}\right) A_{gg\to \chi}(\omega)\right].\end{aligned} \tag{11}$$

In Leading Log approximation one finds:

$$\begin{aligned}A_{g\to gg}(\omega) &= 2\, C_A \left[\frac{11}{12} + \frac{1}{\omega} - \frac{2}{\omega+1} + \frac{1}{\omega+2} - \frac{1}{\omega+3} - S(\omega)\right] \\ A_{gg\to g}(\omega) &= \frac{\pi^2 \rho}{\omega}\, A_{g\to gg}(\omega) \\ A_{g\to g\chi}(\omega) &= \frac{1}{4}\left[\frac{3}{2} - \frac{1}{\omega+1} - \frac{1}{\omega+2} - 2 S(\omega)\right] \\ A_{gg\to \chi}(\omega) &= \frac{\pi^2 \rho}{2\omega}\left[\frac{1}{\omega+1} - \frac{2}{\omega+2} - \frac{2}{\omega+3}\right].\end{aligned} \tag{12}$$

Here $C_A = N_c$, $\rho = O(1)$ is a parameter, $S(\omega) = \psi(\omega+1) - \psi(1)$ with the Psi (Digamma) function $\psi(z) = d[\ln\Gamma(z)]/dz$ and $-\psi(1) = \gamma_E = 0.5772$ the Euler constant. Fig. 2 shows $\gamma(\omega, k_\perp^2)$ for different values of $k_\perp^2$.

The formal solution of eq. (10) is obtained by complex integration in the $\omega$-plane,

$$x\, g(x, k_\perp^2) = \frac{1}{2\pi i} \int_C d\omega\, x^{-\omega} \exp\left\{\int_{k_\perp^2}^{Q^2} \frac{dk_\perp'^2}{k_\perp'^2}\, \gamma(\omega, k_\perp'^2)\right\} \tag{13}$$

which yields the $x$-distribution of gluons as a function of the evolution variable $k_\perp^2$. For the full anomalous dimension (11) this inversion must be done numerically [6]. However, the main features of the evolution of the gluon distribution in the presence of the $\chi$-field become already evident when making some simple analytic estimates. For simplicity I will divide the kinematic range of the $k_\perp^2$-evolution into two distinct domains: (i) $Q^2 \geq k_\perp^2 > Q_0^2$ and $k_\perp^2/\Lambda^2 \gg 1$ so that $\lambda_\chi \simeq 0$ and $\partial\lambda_\chi/\partial \ln k_\perp^2 \simeq 0$, (ii) $Q_0^2 \geq k_\perp^2 \geq \Lambda^2$ and $k_\perp^2/\Lambda^2 \simeq 1$ with $\lambda_\chi \to (4\pi)^{-1}$ and $\partial\lambda_\chi/\partial \ln k_\perp^2 > 0$. In this case, the gluon distribution $g(x, k_\perp^2)$ can be evaluated analytically [6] which yields for $k_\perp^2 \gtrsim Q_0^2$ the well known Leading Log solution of QCD [14], but for $k_\perp^2 < Q_0^2$ a substantial suppression at small $x$ is found to set in, reflecting the "condensation" of gluons into the confining background field $\chi$. The corresponding $k_\perp^2$-dependence of the total gluon multiplicity $N_g(k_\perp^2) = \int_0^1 dx\, g(x, k_\perp^2)$ can also be estimated in the above approximation with the result:

$$(i) \quad N_g(k_\perp^2) = N_g(Q^2) \left(\frac{\ln(Q^2/\Lambda^2)}{\ln(k_\perp^2/\Lambda^2)}\right)^{-1/4} \frac{\exp\left[2\sqrt{(C_A/\pi b)\ln(Q^2/\Lambda^2)}\right]}{\exp\left[2\sqrt{(C_A/\pi b)\ln(k_\perp^2/\Lambda^2)}\right]} \quad (14)$$

$$(ii) \quad N_g(k_\perp^2) = N_g(Q_0^2) \exp\left[-\frac{\rho}{2}\frac{\Lambda^2}{k_\perp^2}\frac{1+[1+\ln(k_\perp^2/Q_0^2)]^2 - 2k_\perp^2/Q_0^2}{\ln(Q_0^2/\Lambda^2)}\right]. \quad (15)$$

In the region (i) $Q^2 \geq k_\perp^2 > Q_0^2$, the multiplicity coincides with the QCD result [14], characterized by a logarithmic growth as the gap between the hard scale $Q^2$ and $k_\perp^2$ increases. On the other hand, in the region (ii) $Q_0^2 \geq k_\perp^2 > \Lambda^2$, the exponent is always negative so that the number of gluons rapidly decreases and vanishes at $k_\perp^2 = 0$, ensuring that no gluons and therefore no color fluctuations exist at distances $R \gtrsim \Lambda^{-1}$. This behaviour is evident in Fig. 3.

Returning to eqs. (6)-(8), the coupled system of equations can now be solved numerically by utilizing the definition (9) together with the solution for the gluon distribution (10) and (13). Since $\chi$ is taken to be equal the mean field $\bar{\chi}$, the multiplication of eq. (7) by $\langle P|$ and $|P\rangle$ from the left and right, respectively, affects only the $F_{\mu\nu}F^{\mu\nu}$ term which gives the gluon distribution by virtue of (9). With the solution of the gluon distribution inserted, the remaining task is to solve eq. (7) for the $\chi$-field. An interesting phenomenological application [15] is to calculate the string constant, which characterizes the linearly rising potential between the $q\bar{q}$-pair due to the gluon interactions. Considering the fragmentation of a heavy $q\bar{q}$ pair, an *adiabatic* (Born-Oppenheimer) treatment is a reasonable approximation, because one can view the $i^{th}$ gluon as being emitted from the $q\bar{q}$ pair plus gluons $g_1, g_2, \ldots, g_{i-1}$, with the spatial coordinates of these "sources" being frozen during the emission of the gluon $i$ [16]. Using the separation $R_{q\bar{q}}$ of the receding $q\bar{q}$ as a measure of the typical gluon transverse momenta $k_\perp^2 \approx R_{q\bar{q}}^{-2}$ and minimizing the energy per unit length of this system, one obtains in the adiabatic approximation for each gluon configuration at a given $R_{q\bar{q}}$ the form of $\chi$ and the string tension $t$. The total energy per unit length $t$ receives contributions from both the $\chi$-field in the low energy regime and the gluon field in the high energy domain, which on account of (5) is given by [6],

$$t \equiv \frac{E}{R_{q\bar{q}}} = \int dA \left[\frac{1}{2}|\nabla\chi|^2 + V(\chi)\right] + \int dA\, \kappa(\chi) \langle P|\frac{\beta(\alpha_s)}{4\alpha_s} F_{\mu\nu,a}F_a^{\mu\nu}|P\rangle, \quad (16)$$

where $A$ is the cross-sectional area of the flux tube of the gluons between the $q\bar{q}$, perpen-dicular to $R_{q\bar{q}}$. In one loop order $\beta = -b\alpha_s^2$ with $b = 33/(12\pi)$ for $N_f = 0$. Using eq. (9), assuming axial symmetry, and minimizing $t$ with respect to $\chi$ yields following non-linear integro-differential equation ($r$ is the radial coordinate perpendicular to $R_{q\bar{q}}$):

$$-\left[\frac{d^2}{dr^2} + \frac{1}{r}\frac{d}{dr}\right]\chi(r) + \frac{\partial V(\chi)}{\partial \chi} + \frac{b\,\alpha_s}{2}(P^+)^2 \int d^2k_\perp \int dx\, x\, g(x, k_\perp^2) \int dr\, r\, \kappa(\chi) = 0. \tag{17}$$

This equation is to be solved subject to the constraint [15] that the total color electric flux through a plane between the $q$ and $\bar{q}$ equals the color charge on one of them, and the boundary conditions $\chi'(0) = 0, \chi(\infty) = \chi_0$. The solutions for $t$, $\chi$ and $\kappa$ are shown in Fig. 4 for the reasonable parameter choice $\chi_0 = f_\pi$ and bag constant $B = (150 \text{ MeV})^4$ [6]. As the quarks separate with increasing $R_{q\bar{q}}$, the gluons first multiply which results in a growing string tension, but then gluon condensation sets in, yielding a saturating behaviour of $t$ (Fig. 4a). For $R_{q\bar{q}} \approx 1$ fm the gluon field is completely confined within a flux tube of radius $r \simeq 1$ fm (Fig. 4b).

In perspective, the Lagrangian approach presented here to describe the confinement dynamics of high energy partons in the framework of effective QCD field theory has the potential to be extended to a systematic quantum description of hadronization. With the inclusion of quark degrees and quantum fluctuations of the $\chi$- and $U$-fields, one could calculate e.g. the mass spectrum of glueball and meson excitations as physical hadrons. Ultimately one would like to address the microscopic dynamics in full 6-dimensional phase-space, with explicit inclusion of the color flow throughout the evolution. The possible applications are manifold. One particular interest is the expected (non-)equilibrium QCD phase transition in high energy systems as in heavy ion collisions or the early universe, an issue which could be addressed along the lines of Campbell, Ellis and Olive [7].

## Acknowledgements

I would like to thank J. Ellis and H.-T. Elze for valuable suggestions.

**FIGURE CAPTIONS**

**Figure 1:** Diagrammatic representation of the two-point Greens function of gluons, including both the gluon (self) interactions and the effective interaction with the confining background field $\chi$ (indicated by the dashed lines). This gluon propagator describes the evolution of a gluon from a chosen cascade branch in $x$ and $k^2$, starting from $x_0$ and $Q^2$.

**Figure 2:** The anomalous dimension $\gamma(\omega, k_\perp^2)$ of eq. (11) versus $\omega$ for different values of $k_\perp^2$.

**Figure 3:** Evolution of the gluon multiplicity $N_g(k_\perp^2)$ from $Q^2$ down to $\Lambda^2$. At first the gluons multiply, but at $k^2 \approx Q_0^2$ a rapid condensation sets which is complete at $\Lambda^2$ ($Q_0 = 1$ GeV, $\Lambda = 0.23$ GeV).

**Figure 4:** a) String tension $t$ versus separation $R_{q\bar{q}}$ of the $q\bar{q}$ pair, and b) the solutions for $\chi$ and $\kappa$ at $R_{q\bar{q}} = 1$ fm versus $r$ which is the radial coordinate perpendicular to $R_{q\bar{q}}$ ($\chi_0 = f_\pi$ and $B^{1/4} = 150\,\text{MeV}$).

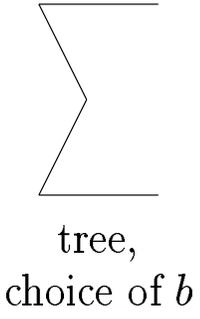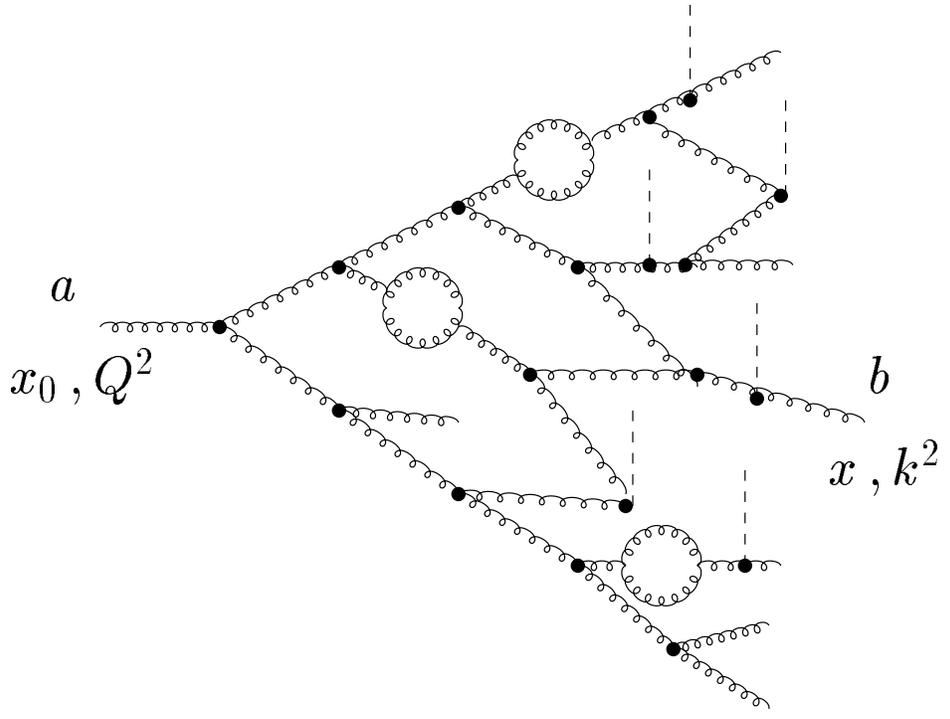

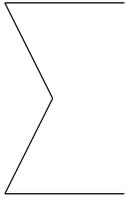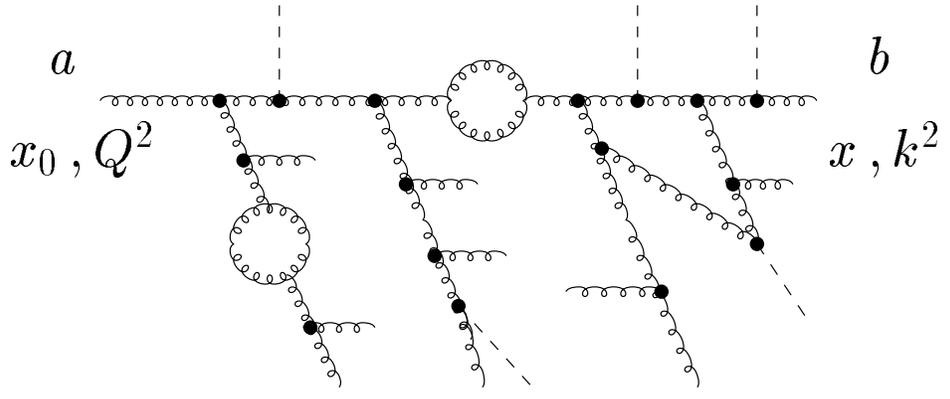

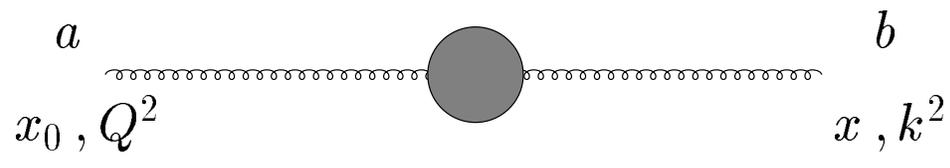

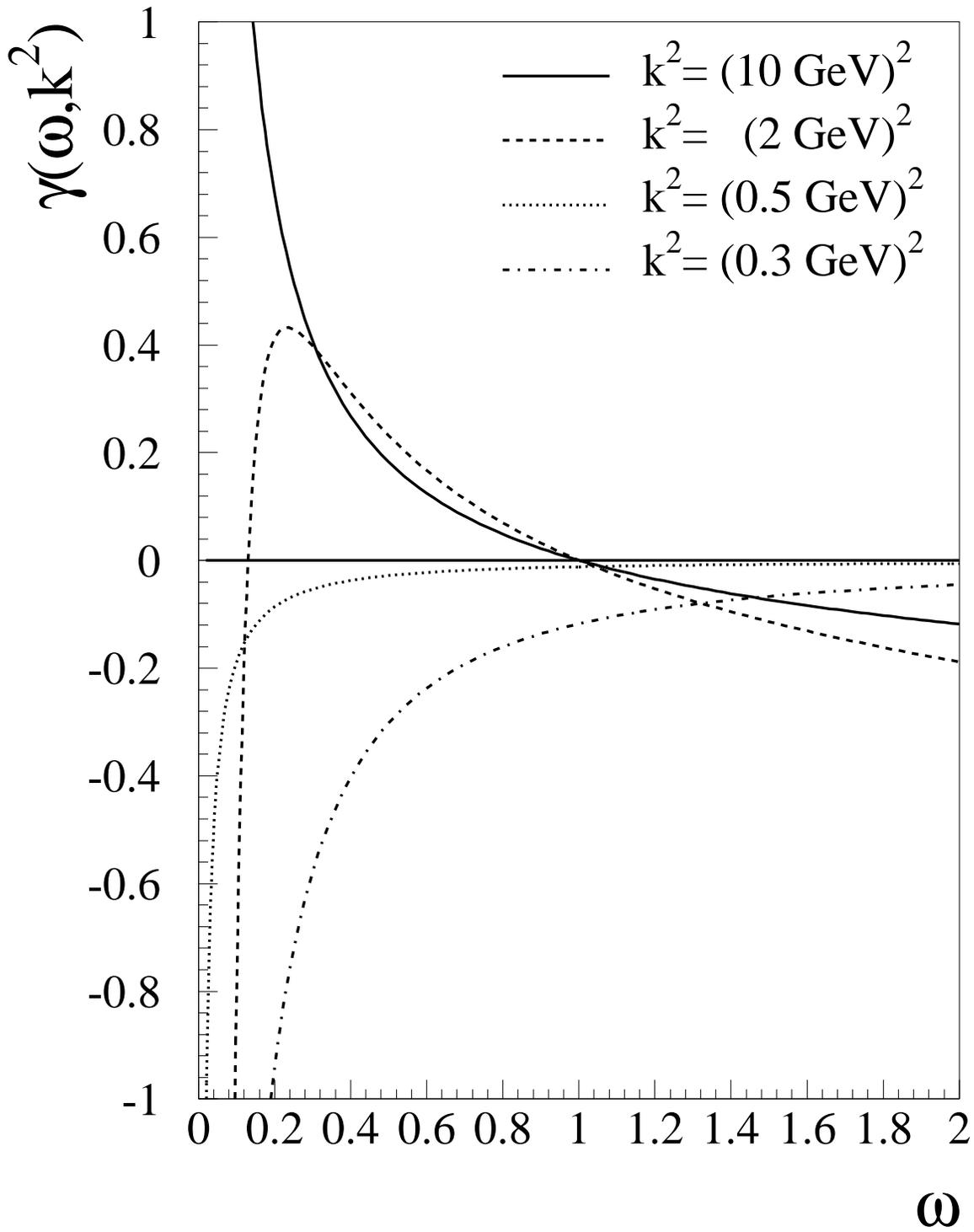

Fig. 2

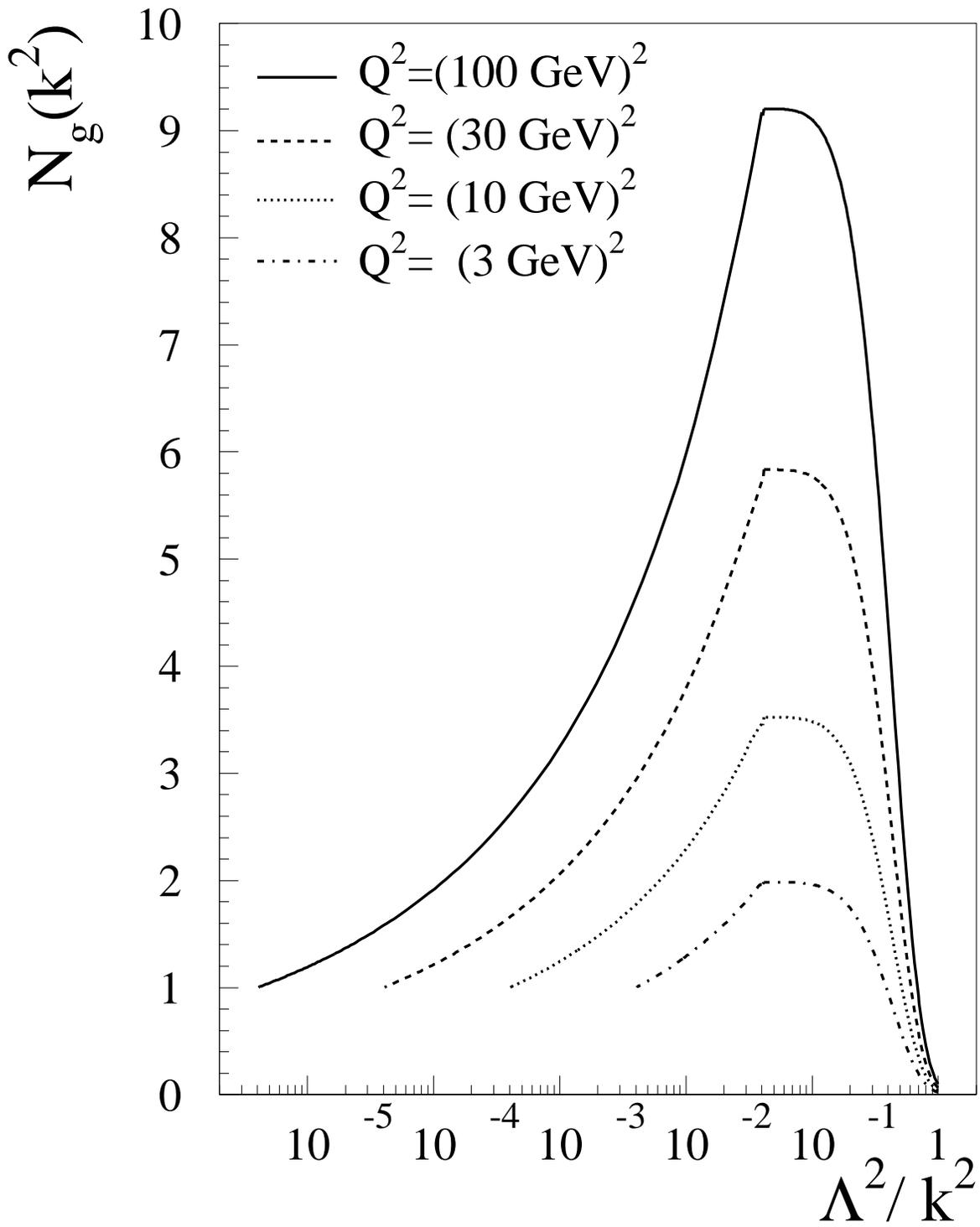

Fig. 4

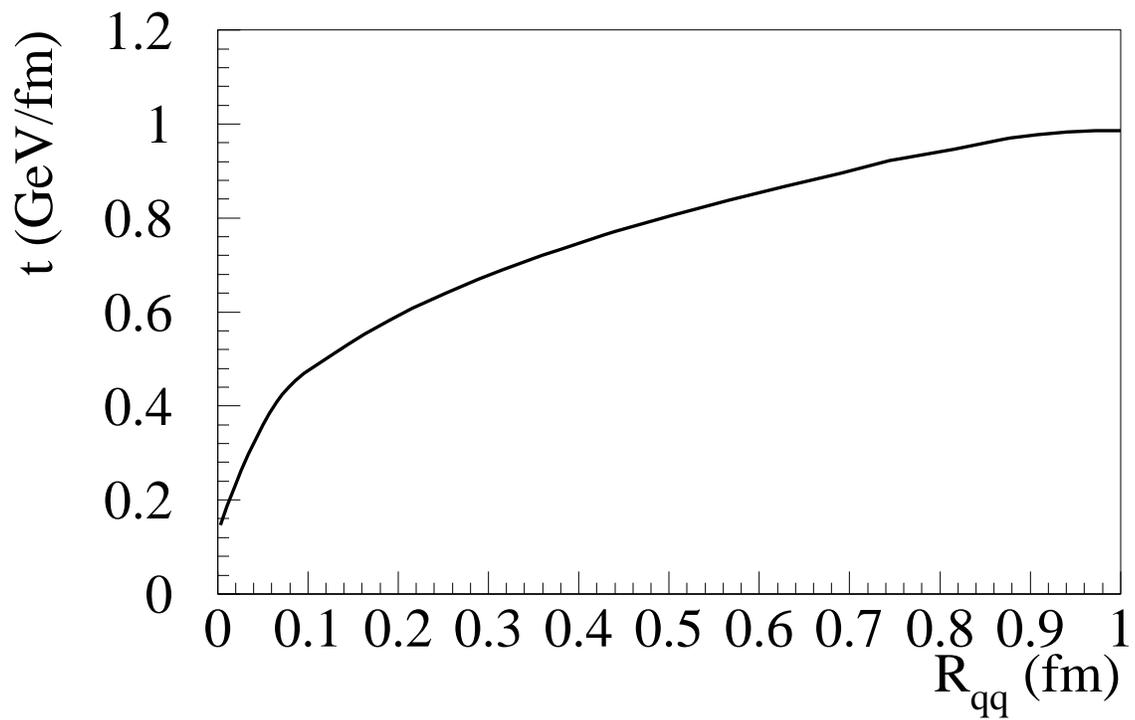

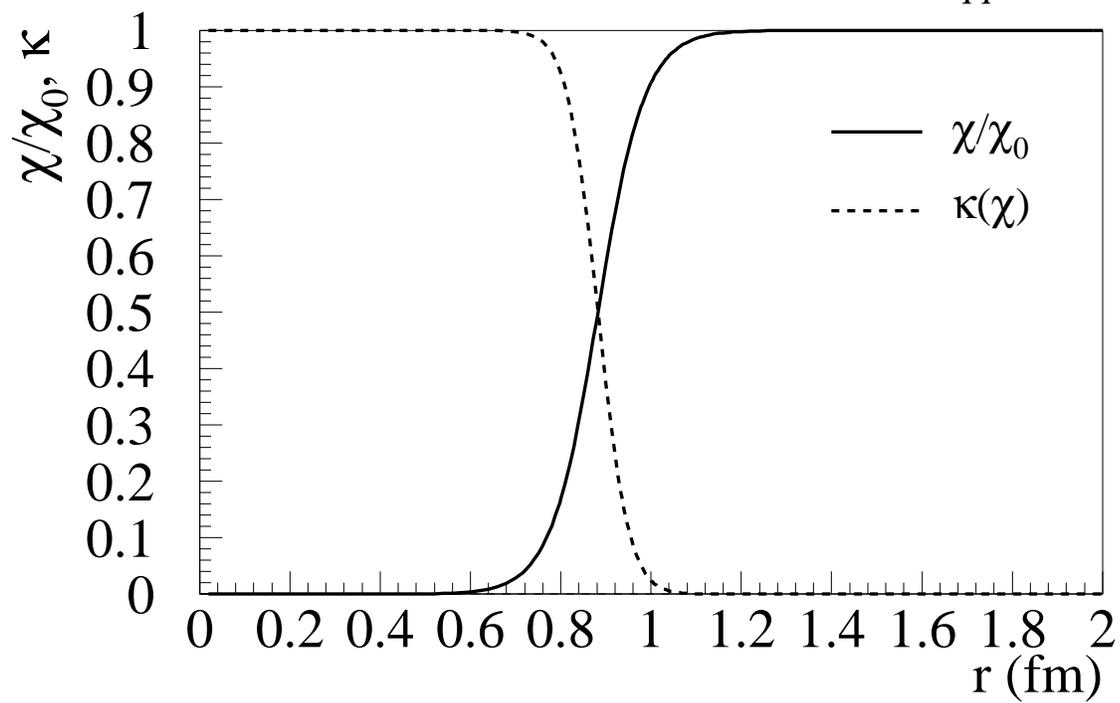